\documentstyle[graphicx]{aipproc}

\newcommand{\muu}{\mbox{$\mu$}} 
\newcommand{\muw}{\mbox{$\mu_{\rm W}$}}
\newcommand{\mui}{\mbox{$\mu_{\rm I}$}}
\newcommand{\sig}{\mbox{$\sigma$}}
\newcommand{\sigw}{\mbox{$\sigma_{\rm W}$}}
\newcommand{\sigi}{\mbox{$\sigma_{\rm I}$}}

\title{Lognormal properties of SGR 1806-20\\ and the Possibility of a Quiescent Population of other SGR sources}

\author{Brian McBreen and Kevin J. Hurley}
\address{Physics Dept., University College, Dublin 4, Ireland.}

\begin{document}

\maketitle

\begin{abstract}
  
  Monte Carlo simulations of
  long SGR event sequences based on lognormal distributions with a
  range of time intervals and intensity distribution parameters have
  been investigated. The main conclusions are that the majority of 
  SGRs with
  properties similar to SGR 1806-20 have been detected but SGRs with
  mean waiting times much longer than SGR 1806-20 remain to be
  discovered. A large decrease in the probability for detection of an
SGR source results from a relatively small increase in the distribution
parameters obtained for SGR 1806-20.  A new breed of experiments with very long observation
  times are required to search for this type of source.

\end{abstract}

\section*{Introduction}

The lognormal properties of the soft repeater SGR1806-20 have been
previously reported\cite{sg-p07:cheng.b:95-nat-382-518,%
sg-p07:hurley.kj:95-apss-231-81,%
sg-p07:hurley.kj:94-aa-288-l49}. In particular, both the time interval
between repeater events and the luminosity function of the source were
fit with lognormal distributions\cite{sg-p07:aitchison.j:57-b-tld}. This
analysis used the data-base of 111 events detected by the
International Cometary Explorer (ICE)
mission\cite{sg-p07:laros.jg:87-apj-320-l111}.

While the present number of events observed from the other three
sources \cite{sg-p07:kouveliotou.c:93-nat-362-728,sg-p07:norris.jp:91-apj-366-240,%
sg-p07:new-sgr-source} does not allow any detailed
analysis, the intervals between successive events of SGR 0526-66
\cite{sg-p07:golenetskii.sv:87-sal-13-3-166} is also suggestive of
lognormal behaviour.

The relationship between the number of active (i.e observable) sources
and the true number of SGRs in the galaxy is one which is the subject
of some debate \cite{sg-p07:hurley.kc:94-apj-423-709,%
sg-p07:kouveliotou.c:92-apj-392-179}. If the time intervals between SGR
events proves to be lognormal with a wide range of means and variances
then there may be long quiescent periods where the source could be
undetectable, leading to an underestimate of the population.

\section*{Simulations}

Monte Carlo simulations of SGR event sequences with a variety of
distribution parameters were generated. The simulations were performed
using Octave 2.0.5 for UNIX. Randomly generated standard normal
variates, $X$, were transformed to lognormal variates using the
relationship $Y=e^{\sigma \! X+\mu}$ where $Y$ is lognormally
distributed with parameters \muu\ and \sig. The effect of the
detector on the observations was simulated by rejecting events below a
preset intensity level to mimic a threshold and by rejecting at
random some predefined fraction of events to mimic a less than 100\%
live time.

\begin{figure}[tb]
  \includegraphics[angle=270,width=\textwidth]{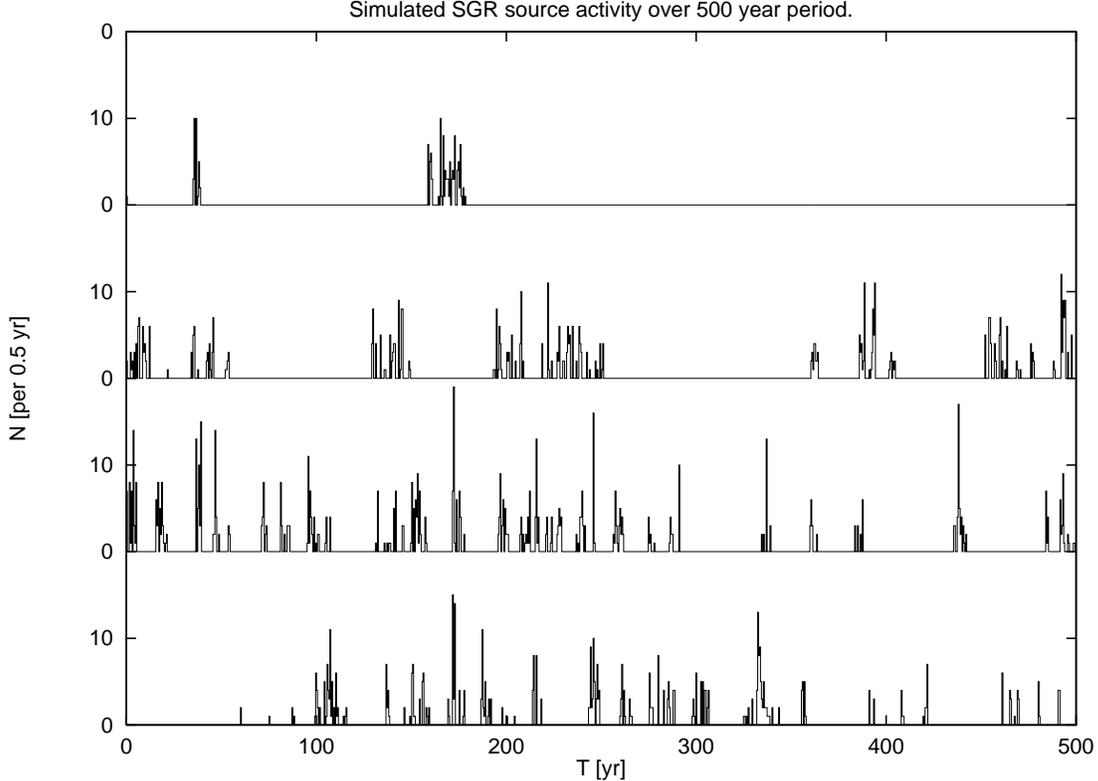}
  
  \caption[]{\label{sg-p07:fig:gen-seq}
    Four separate Monte Carlo simulations of duration 500 years of SGR1806-20
    activity, generated from lognormal distributions for recurrence
    intervals and intensities with parameters as determined from
    SGR1806-20 \protect{\cite{sg-p07:hurley.kj:94-aa-288-l49}}. 
    The detector model is
    described in the text. The long gaps in activity of the source
    are the contributions from the tail of this highly skewed
    distribution.}
\end{figure}

\begin{figure}[tb]
\includegraphics[bb= 35 35 495 745,clip, angle=90, width=\textwidth]{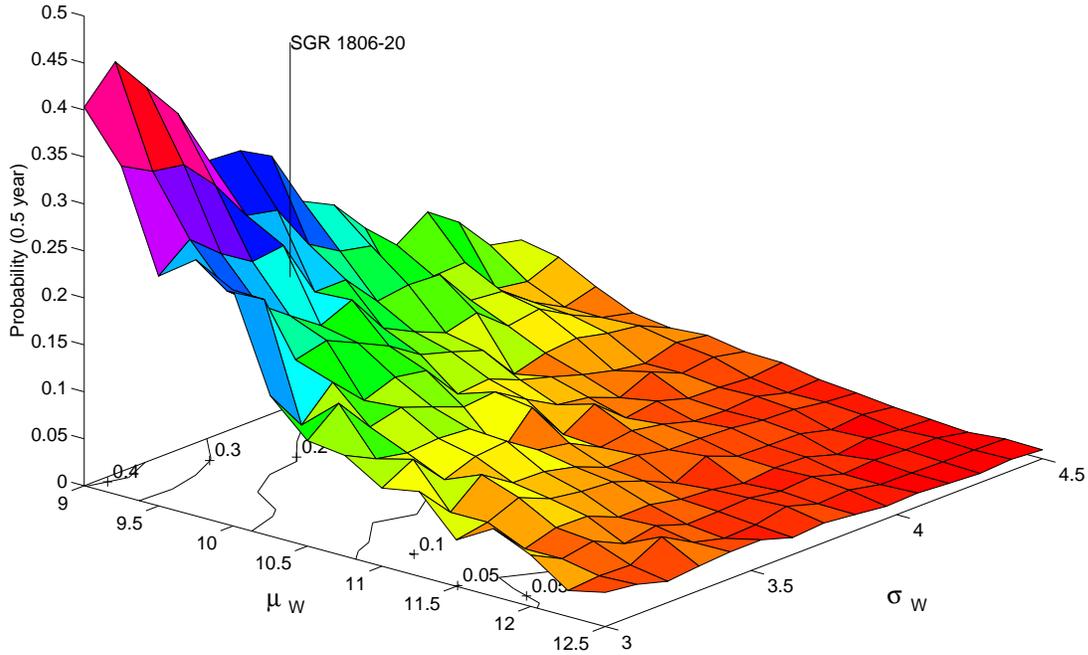}
\caption{\label{sg-p07:fig:surf-w}%
  This surface plot shows how the probability of observing an SGR
  source varies with the parameters of the waiting time
  distribution(\muw and \sigw). The probability given is per 0.5
  years. The contour lines show the drop in probability with \muw\ \, and \sigw. The indicated cell contains the value for SGR 1806-20 with
  $\muw=9.6$ and $\sigw=3.4$.}
\end{figure}

\begin{figure}[tb]
\includegraphics[bb= 0 0 500 790, clip, angle=90, width=\textwidth]{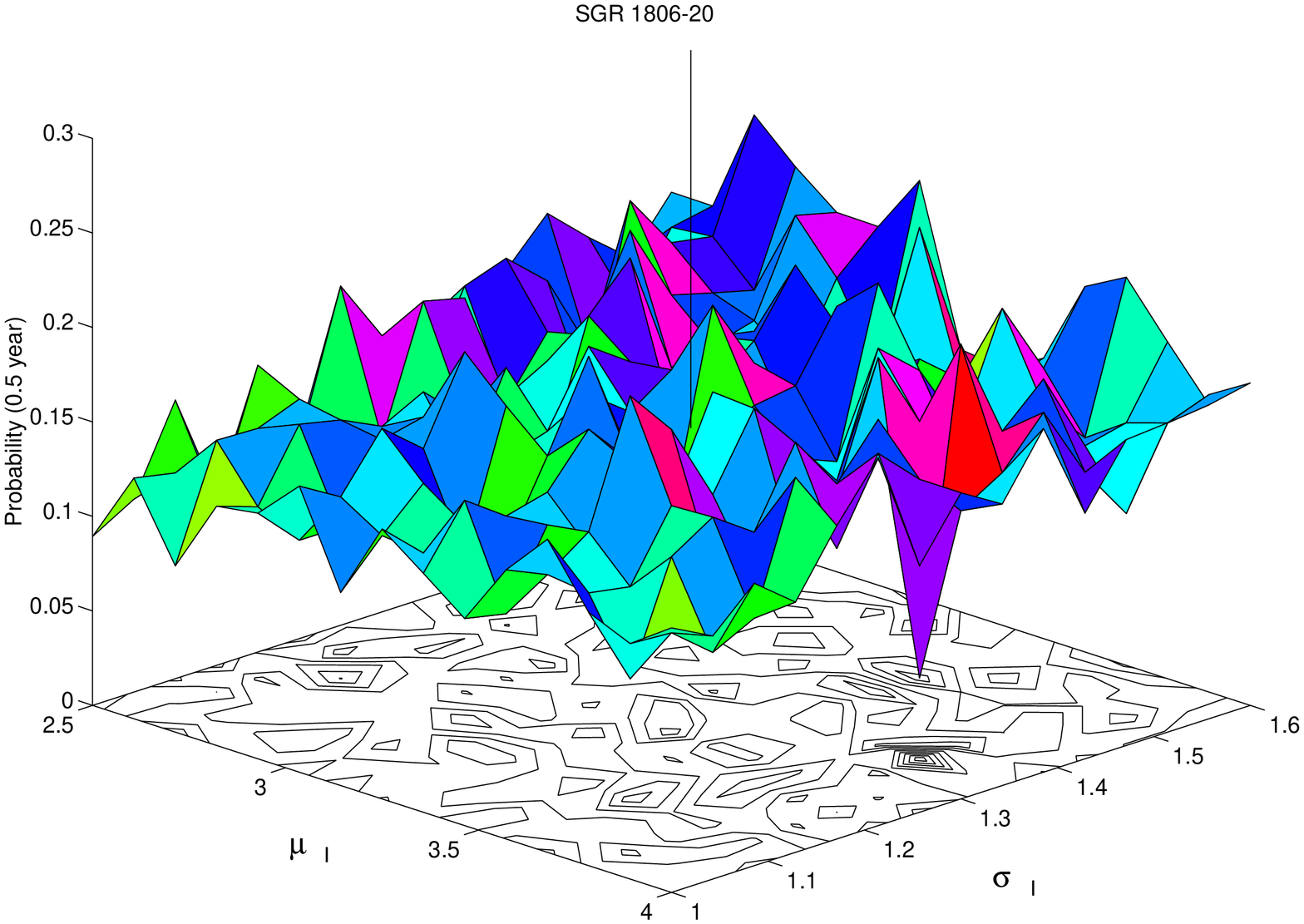}
\caption{\label{sg-p07:fig:surf-i}%
  This surface plot shows how the probability of observing an SGR
  source varies with the parameters of the intensity distribution(\mui
  and \sigi). The probability given is per 0.5 years. The 
  indicated cell contains the value for SGR 1806-20 with $\mui=3.3$ and
  $\sigi=1.3$.}
\end{figure}

The parameters of the two lognormal density function which were fit to
the distribution of recurrence intervals and intensity for SGR1806-20
were respectively $\muw=9.64$, $\sigw=3.44$ and $\mui=3.3$,
  $\sigi=1.3$ \cite{sg-p07:hurley.kj:94-aa-288-l49}.  To investigate how
the probability of observation with a given time interval depended on
the the parameters for the waiting times (\muw\ and \sigw) and the
parameters for the event intensity (\mui\ and \sigi), two large
simulations were performed. In the first, the waiting time parameters
were varied over a small range relative to the
SGR1806-20 values (while the intensity distribution was fixed), forming
a grid of $14\times14$ cells, each with a different value
for \muw\ and \sigw\ but the same value for \mui\ and \sigi. For each cell on the grid ten 500 year simulations were
performed (as for the sequences produced in
Figure~\ref{sg-p07:fig:gen-seq}) and the mean number of events per 0.5
year interval was determined. The second simulation was performed with varying
\mui\ and \sigi (the intensity distribution parameters) but it followed
the same procedure - the region of parameter space to be explored was divided into a
grid and ten 500 year simulations were performed at each grid point
(each grid point having a different \mui\ and \sigi\ but the same \muw\ 
and \sigw). The resulting distributions of probability of observation
for each of the two grids are presented in
Figures~\ref{sg-p07:fig:surf-w} and Figure~\ref{sg-p07:fig:surf-i}.

\section*{Discussion}

Figure~\ref{sg-p07:fig:surf-w} presents a surface plot of the probability
of detecting an SGR source within 0.5 years, given that the event
intervals are distributed with a lognormal probability and the intensity
distribution is fixed to that observed for SGR1806-20. Note that the
parameters for SGR1806-20 ($\mu_W$ and $\sigma_W$) an estimated 
probability of 30\% for observation in 0.5 years and that the probability
falls quickly with increasing \muw\ and \sigw. For $\muw = 11 \approx
1.2 \mu_W^{1806}, \sigw = 4.5 \approx 1.5\sigma_W^{1806}$, which is equivalent to a mean
of $1.5 \times 10^{9} {\rm s}$, the probability of observing the source
in $0.5$ years has fallen to less than 5\%.

In contrast, Figure~\ref{sg-p07:fig:surf-i} shows a much flatter more
uniform behaviour over a similar variation in the intensity
distribution parameters.  Note that the parameters for SGR1806-20
again give us an estimated probability of 30\% for observation in the
0.5 years of observation.  The grainy nature of the surface is due to
the much lower variation in probability across the range of
intensities chosen (from 50\% of SGR1806-20 to 300\% for each
distribution parameter).

The lognormal distribution arises in statistical processes whose
completion depend on a product of probabilities, arising from a
combination of independent events
\cite{sg-p07:montroll.ew:82-pnasusa-79-3380}.  Lognormal statistics have
previously been used in connection with gamma-ray bursts
\cite{sg-p07:brock.m:94-aip-307-672,sg-p07:mcbreen.b:94-mn-271-662}. In the
context of this investigation the physical significance of this
statistical behaviour may lie in the connection between SGRs and
neutron stars where a similar statistical analysis was presented for
the micro\-glitches from the Vela pulsar. The time separation and the
intensity of these small ($\mid\Delta\nu/\nu\mid\sim 10^{-9}$) frequency adjustments were both compatible with lognormal
distributions, and there was no correlation between waiting time and
intensity, just as observed with SGR1806-20
\cite{sg-p07:laros.jg:87-apj-320-l111}. This result, combined with the
identification of X-ray point sources
\cite{sg-p07:murakami.t:94-nat-368-127,sg-p07:rothschild.re:94-nat-368-432}
embedded in plerion-powered SNR \cite{sg-p07:kulkarni.sr:93-nat-365-33}
as counterparts to the SGR sources, suggests structural adjustments in
neutron stars may be the cause of SGRs.

\section*{Conclusion}

The activity of sources with mean recurrence times similar to or much
longer than SGR 1806-20 was investigated using Monte Carlo
simulations. The results of the simulations indicate that there could
exist a significant population of SGRs with mean waiting times
considerably longer than SGR1806-20 that remain undiscovered.
Structural adjustments in neutron stars may be responsible for this
behaviour. Finally, a new breed of experiments with very long
observation times will be required to search for this type of source.


%

\end{document}